\documentclass{article}
\usepackage{spconf,amsmath,graphicx}
\usepackage{xcolor}
\usepackage{amsmath,bbm}
\usepackage{amssymb}
\usepackage{booktabs}
\usepackage{multirow}
\usepackage{cite}
\usepackage{hyperref}
\usepackage[mathscr]{euscript}
\usepackage{fancyhdr}

\DeclareMathAlphabet\mathbfcal{OMS}{cmsy}{b}{n}

\title{Latent-space Scalability for Multi-task  Collaborative Intelligence}
\name{Hyomin Choi and Ivan V. Baji\'{c}}
\address{School of Engineering Science, Simon Fraser University, Burnaby, BC, Canada}
\begin{document}
%
\maketitle

\thispagestyle{empty}
\setlength{\headheight}{22.5pt}
\renewcommand{\headrulewidth}{0.0pt}
\thispagestyle{fancy}
\lhead{}
\chead{Copyright \copyright 2021 IEEE. Personal use of this material is permitted. However, permission to use this material for any other purposes must be obtained from the IEEE by sending an email to pubs-permissions@ieee.org.}
\rhead{}
\lfoot{}
\cfoot{}
\rfoot{}

\begin{abstract}
We investigate latent-space scalability for multi-task collabo-rative intelligence, where one of the tasks is object detection and the other is input reconstruction. In our proposed approach, part of the latent space can be selectively decoded to support object detection while the remainder can be decoded when input reconstruction is needed. Such an approach allows reduced computational resources when only object detection is required, and this can be achieved without reconstructing input pixels. By varying the scaling factors of various terms in the training loss function, the system can be trained to achieve various trade-offs between object detection accuracy and input reconstruction quality. Experiments are conducted to demonstrate the adjustable system performance on 
the two tasks compared to the relevant benchmarks.

\end{abstract}
\section{Introduction}
\label{sec:intro}

Rapid deployment of artificial intelligence (AI)-enabled applications is putting a strain on computational resources across a number of systems, from handheld devices to large-scale cloud computing systems. Recent studies~\cite{Kang2017, eshratifar2019jointdnn} have established the concept of \emph{collaborative intelligence} (CI) as one way to address such challenges, by splitting an AI model (e.g., a deep neural network, DNN) between the edge and the cloud. 
In such a framework, intermediate features, produced by the model's front-end, are sent from the edge to the cloud. 
Hence, compression of intermediate features has become a topic of interest. Related standardization activities include Video Coding for Machines (VCM)~\cite{vcm_call_for_evidence} and JPEG-AI~\cite{JPEG-AI_use_cases}. 

For example,~\cite{dfc_for_collab_object_detection, eshratifar2019towards, eshratifar2019bottlenet,   Choi_BaF_2020, Cohen_quantcode_ICME2020} have demonstrated that coding intermediate features can lead to significant compression gains, with a minimal loss in task accuracy. 
These studies were based on off-the-shelf single-task DNN models. In our earlier work~\cite{Choi2018NearLosslessDF}, a multi-task CI model was developed that supports object detection and input reconstruction, using near-lossless coding of intermediate features. 
Related approaches~\cite{torfason2018towards, saeed_multi_task_learning}, utilizing lossy feature compression, were presented for different multi-task models. Unlike these methods, where a single feature tensor is coded to support multiple back-end tasks, recent proposals~\cite{duan2020video, hu2020towards} focus on scalable coding to support multiple tasks. For example,~\cite{hu2020towards} presented a scalable coding approach that supports facial landmark detection and generative input face reconstruction. While the generative decoder works well for face reconstruction, it might be less successful in reconstructing non-face details of the input image.


In this paper, we present a CI system that uses \emph{latent-space scalability} to support object detection and input image reconstruction. Specifically, a part of the latent space (base layer) is dedicated to object detection (base task) while the entire latent space is utilized for input reconstruction. The portion of the latent space that is not used for the base task can be interpreted as an enhancement layer. Such representation can also be used for other multi-task models (i.e., the base task could be something other than object detection) and allows for efficient, scalable learnt representation of the input.  

Section~\ref{sec:prev_work} briefly reviews related approaches to intermediate feature compression. 
The proposed method is described in Section~\ref{sec:proposed}. Experimental results are presented in Section~\ref{sec:experiments}, followed by conclusions in Section~\ref{sec:conclusion}.

\section{Related work}
\label{sec:prev_work}
Early approaches to feature compression~\cite{eshratifar2019bottlenet, eshratifar2019towards, dfc_for_collab_object_detection, Choi_BaF_2020, Cohen_quantcode_ICME2020} focused on coding a single feature tensor from a single-task DNN, with tasks being image classification~\cite{eshratifar2019bottlenet, eshratifar2019towards} or object detection~\cite{dfc_for_collab_object_detection}. 
A popular approach for coding feature tensors in these works was to tile the tensor into an image, apply pre-quantization (say, to 8 bits per tensor element), and then use a conventional image codec for compression. In order to further improve the tensor coding efficiency,~\cite{Choi_BaF_2020, Cohen_quantcode_ICME2020} proposed additional methods such as tensor channel prediction and data clipping. 

Since multiple tasks are often required in image/video analysis~\cite{zhang2016joint, duan2020video}, another group of methods has focused on feature compression for multi-task DNNs~\cite{torfason2018towards, Choi2018NearLosslessDF, saeed_multi_task_learning}. Although these works validated the idea that multi-task analytics are possible from a single compressed feature tensor, no further study was made as to how to efficiently organize the latent space for multiple tasks. In particular, in these approaches, reconstruction of the whole tensor is needed to accomplish any task.  
Most recently,~\cite{hu2020towards} proposed a scalable feature representation for coding face images. Specifically, edge maps needed for facial landmark detection form the base layer, while additional color information forms the enhancement layer. Facial landmark detection can be accomplished from the base-layer information only, while facial image can be reconstructed using both base and enhancement layer, by a generative decoder. While the main idea in~\cite{hu2020towards} is very appealing, it is not clear how this approach can be extended to more general (e.g., non-face) image coding scenarios. 

The approach presented in this paper builds on the idea of scalable latent-space representation, and is more widely-applicable than~\cite{hu2020towards}. In particular, it can accommodate generic learnt features and an arbitrary base task. For concreteness, our experiments are carried out on a model whose base task is object detection, but it should be noted that a similar methodology could be applied to another base task, such as image classification, object segmentation, etc. 


\section{Proposed method}
\label{sec:proposed}

\begin{figure}[t]
    \centering
    \begin{minipage}[b]{0.6\linewidth}
    \centering
    \includegraphics[width=\textwidth]{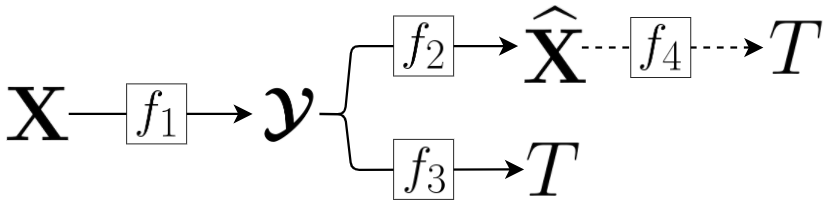}
    \end{minipage}
\caption{Markov chain model for our CI system.}
\label{fig:intuitive_method}
\end{figure}

\subsection{Motivation}
A Markov chain model for the CI system studied in this paper is shown in Fig.~\ref{fig:intuitive_method}. Input image $\mathbf{X}$ is processed by the edge sub-model $f_1$, producing features $\mathbfcal{Y}$. At the cloud side, from the features $\mathbfcal{Y}$, sub-model $f_2$ reconstructs an approximation $\widehat{\mathbf{X}}$ to the input image $\mathbf{X}$, while sub-model $f_3$ performs object detection, producing a collection $T$ of bounding boxes and object classes. 

Processing chain $\mathbf{X} \to \mathbfcal{Y} \to \widehat{\mathbf{X}}$ acts as an end-to-end image codec. Note that object detection can also be performed on the decoded image $\widehat{\mathbf{X}}$, using an off-the-shelf object detector such as YOLO~\cite{Redmon2018_yolov3} or SSD~\cite{liu2016ssd}, which is indicated by $f_4$ in Fig.~\ref{fig:intuitive_method}. In fact, such object detection from decoded images (rather than raw images) is common practice, because object detection datasets such as COCO~\cite{COCO} and ImageNet~\cite{imagenet_cvpr09} contain JPEG-compressed images rather than raw images. Applying data processing inequality~\cite{Cover_Thomas_2006} to the Markov chain $\mathbfcal{Y} \to \widehat{\mathbf{X}} \to T$, we have
\begin{equation}
    I(\mathbfcal{Y};\widehat{\mathbf{X}}) \geq I(\mathbfcal{Y};T),
\label{eq:dpi}
\end{equation}
where $I(\cdot \, ; \, \cdot)$ denotes mutual information~\cite{Cover_Thomas_2006}. This suggests that intermediate features $\mathbfcal{Y}$ carry less information about object detection ($T$) than they do about input reconstruction ($\widehat{\mathbf{X}}$). This observation motivates our approach - we construct the features $\mathbfcal{Y}$ such that only part of $\mathbfcal{Y}$ is used for object detection, while the whole of $\mathbfcal{Y}$ is used for input reconstruction.

\begin{figure}[t]
    \begin{minipage}[b]{\linewidth}
    \centering
    \includegraphics[width=\textwidth]{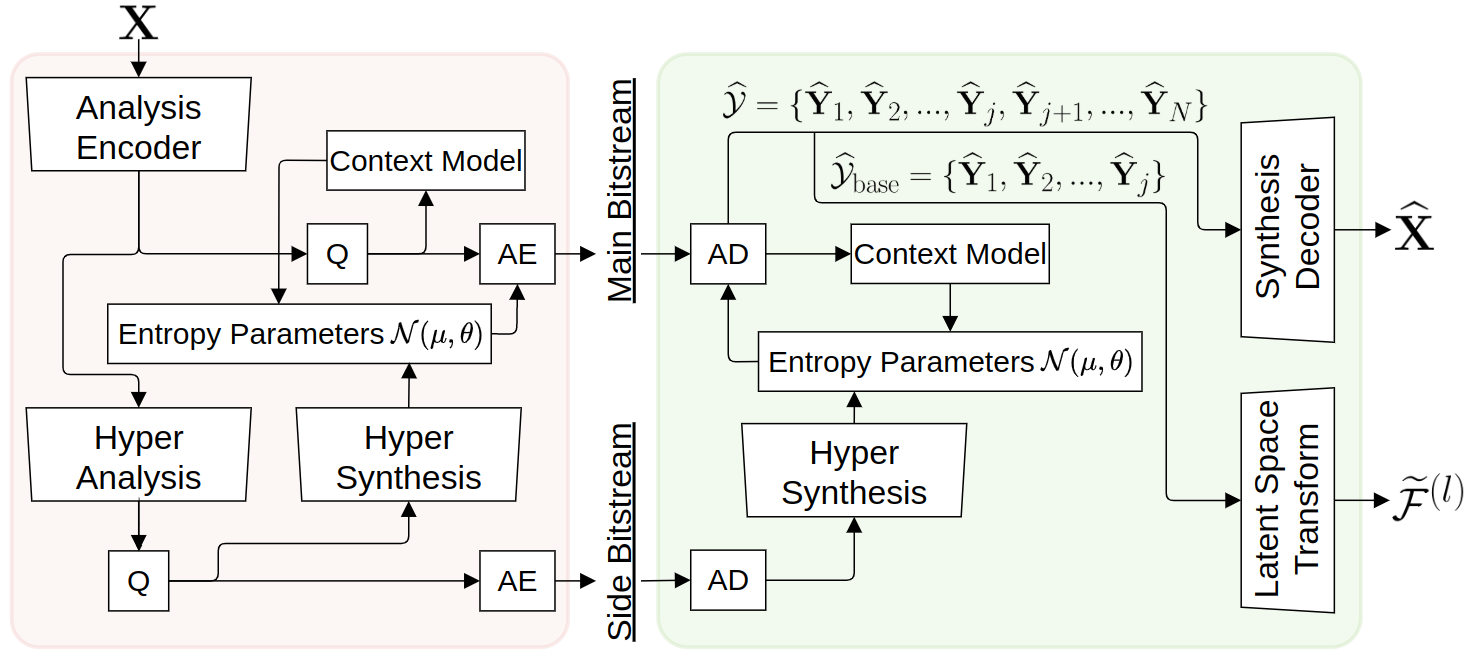}
    \end{minipage}
    \centerline{\hspace{-0.7cm}Edge \hspace{3cm} Cloud}
\caption{Architecture of our CI system. `Q'  represents quantization, `AE'/`AD' represent arithmetic encoder/decoder. Configuration details of `Context Model', `Entropy Parameters' and `Hyper Analysis/Synthesis' follow~\cite{minnen2018joint}, whereas `Analysis Encoder', `Synthesis Decoder' and `Latent Space Transform' are newly proposed for our use case.}
\label{fig:overall_system}
\end{figure}

Fig.~\ref{fig:overall_system} shows the architecture of our CI system. A number of modules in the system are based on~\cite{minnen2018joint}, while the newly proposed modules are discussed in more detail below.

\subsection{Analysis Encoder and Synthesis Decoder}

\begin{figure}[t]
    \centering
    \begin{minipage}[b]{0.49\linewidth}
    \centering
    \includegraphics[width=\textwidth]{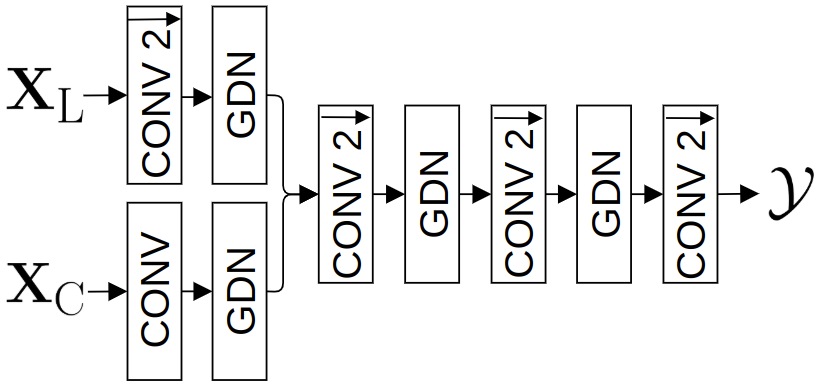}
    \centerline{(a)}\medskip
    \end{minipage}
    \begin{minipage}[b]{0.49\linewidth}
    \centering
    \includegraphics[width=\textwidth]{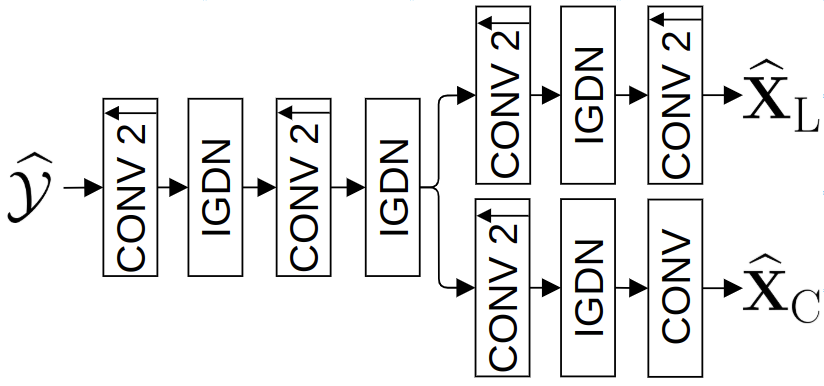}
    \centerline{(b)}\medskip
    \end{minipage}
\caption{Architecture of (a) Analysis Encoder and (b) Synthesis Decoder for YUV420 input/output}
\label{fig:input_output}
\end{figure}

While most end-to-end learnt image compression methods~\cite{balle2018variational, balle2019end, minnen2018joint} are made for RGB input images, we designed our system for YUV420 input format, which is more common in video coding. Specifically, input image $\mathbf{X}$ comprises a luminance channel $\mathbf{X}_{\textup{L}} \in \mathbb{R}^{1\times H \times W}$ and chrominance channels $\mathbf{X}_{\textup{C}} \in \mathbb{R}^{2 \times H/2 \times W/2 }$, where $H \times W$ is the input resolution. 
The architectures of the corresponding Analysis Encoder and Synthesis decoder are shown in Fig.~\ref{fig:input_output}. The Analysis Encoder comprises a number of convolutional (`CONV') layers (with $5 \times 5$ filters) and generalized divisive normalization (GDN)~\cite{balle2015density} layers. Downsampling in the luminance branch is performed by a convolution with stride 2. Synthesis Decoder is a mirror of the Analysis Encoder, with convolutions replaced by transpose convolutions (indicated by $\uparrow$) and GDN layers replaced by inverse GDN (IGDN) layers.
At the output of the Synthesis Decoder, the reconstructed input $\widehat{\mathbf{X}}$ consists of $\widehat{\mathbf{X}}_{\textup{L}}$ and $\widehat{\mathbf{X}}_{\textup{C}}$.

\subsection{Latent-space scalability}
\label{ssec:qlayer}
The latent-space feature tensor in our system is of dimensions $\mathbfcal{Y} \in \mathbb{R}^{N \times H/16 \times W/16}$, consisting of $N=192$ channels: $\mathbfcal{Y}=\{\mathbf{Y}_1, \mathbf{Y}_2, ..., \mathbf{Y}_N\}$. We split this tensor into two parts, $\mathbfcal{Y}_{\textup{base}}=\{\mathbf{Y}_1, \mathbf{Y}_2, ..., \mathbf{Y}_j\}$, representing the base-layer features with $j<N$ channels, and $\mathbfcal{Y}_{\textup{enh}}=\{\mathbf{Y}_{j+1}, \mathbf{Y}_{j+2}, ..., \mathbf{Y}_N\}$, representing the enhancement-layer features with $N-j$ channels. In our experiments, we used $j=128$. At the decoder, if only object detection is required, only $\mathbfcal{Y}_{\textup{base}}$ needs to be reconstructed. If input image reconstruction is required, then the entire $\mathbfcal{Y}$ is reconstructed.

An obvious question is - how do we know that the object detection-related information is concentrated in the first $j$ channels of $\mathbfcal{Y}$? This is achieved by training the entire model in Fig.~\ref{fig:overall_system} from scratch, as explained in Section~\ref{ssec:training}. Through gradient-based updates from various loss terms, the model learns to steer the object detection-relevant information into $\mathbfcal{Y}_{\textup{base}}$, while at the same time learning to reconstruct the input image using the entire $\mathbfcal{Y}$.


\subsection{Latent space transform}
\label{ssec:lst}

\begin{figure}[t]
    \begin{minipage}[b]{\linewidth}
    \centering
    \includegraphics[width=\textwidth]{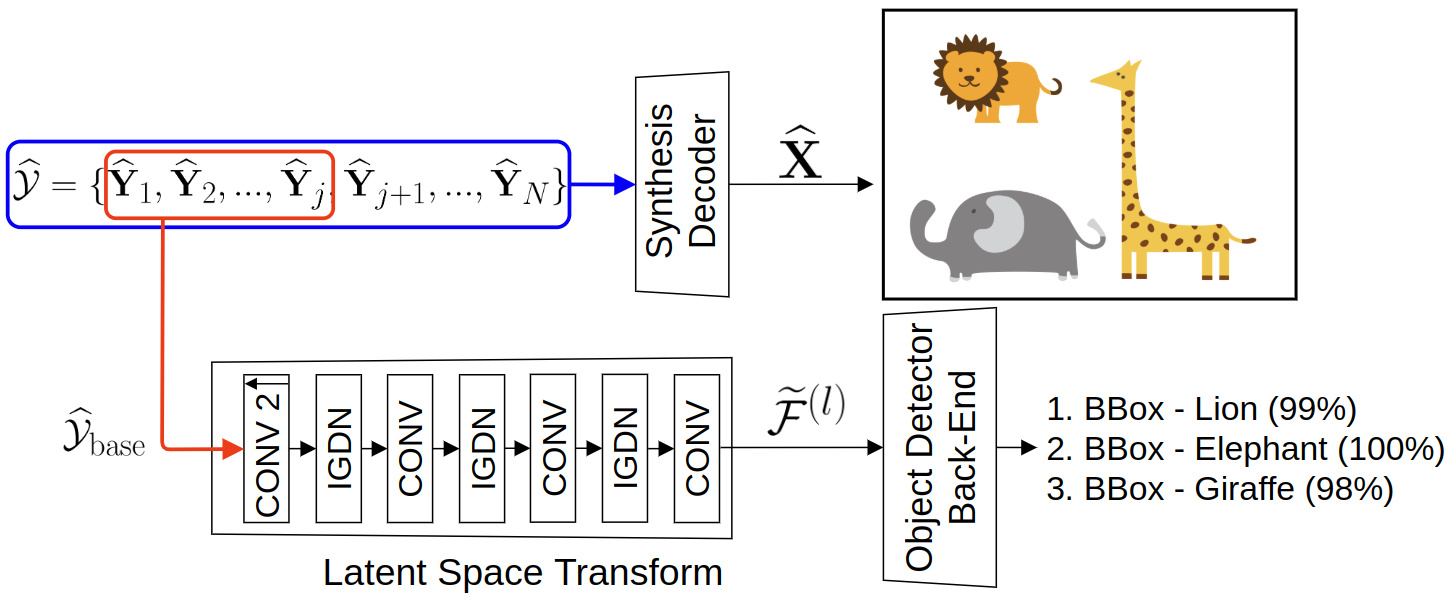}
    \end{minipage}
\caption{Latent space transform to enable object detection from a subset of $\widehat{\mathbfcal{Y}}$} 
\label{fig:latent_transform}
\end{figure}
We use the pre-trained back-end of YOLOv3~\cite{Redmon2018_yolov3} for object detection in our system, specifically the portion from the batch normalization input in layer $l=12$ up to the model output. At this point, YOLOv3 expects a feature tensor $\mathbfcal{F}^{(l)} \in \mathbb{R}^{256 \times H/8 \times W/8}$, whereas our reconstructed base  features are $\widehat{\mathbfcal{Y}}_{\textup{base}} \in \mathbb{R}^{128 \times H/16 \times W/16}$. Hence, a transformation from one latent space to another is needed. The structure of the latent space transform module is shown in Fig.~\ref{fig:latent_transform}; it consists of a transpose convolutional layer, whose purpose is to match the spatial resolution of the target latent space, and and a sequence of IGDN and convolutional layers. At the output, a feature tensor in the target latent space,  $\widetilde{\mathbfcal{F}}^{(l)}$, is produced.
Once $\widetilde{\mathbfcal{F}}^{(l)}$ is computed, it is fed to the batch normalization $\textup{B}^{(l)}$ of layer $l=12$ of YOLOv3, followed by $\texttt{LeakyReLU}$ activation $\sigma(\cdot)$, thus producing the input to layer $l=13$. 

\subsection{Training}
\label{ssec:training}
Our loss function is in the form of a rate-distortion Lagrangian
\begin{equation}
    \mathcal{L} = R + \lambda \cdot D,
    \label{eq:RD_loss}
\end{equation}
where $R$ is the rate estimate, $D$ is the combined distortion for both input reconstruction and object detection, and $\lambda$ is the Lagrange multiplier. Since our coding engine is based on~\cite{minnen2018joint}, the bitstream consists of the main bitstream, encoding latent data, and the side bitstream encoding the hyper-priors. Based on~\cite{minnen2018joint}, rate estimates for these two bitstreams are  
\begin{equation}
R=\underbrace{\mathbb{E}_{x\sim p_x}\left [ -\textup{log}_{2}p_{\hat{y}}(\hat{y}) \right ]}_{\textup{main bitstream}}+\underbrace{\mathbb{E}_{x\sim p_x}\left [ -\textup{log}_{2}p_{\hat{z}}(\hat{z}) \right ]}_{\textup{side bitstream}},
\label{eq:rate_term}
\end{equation}
where $x$ denotes input data, $\hat{y}$ denotes latent data, and $\hat{z}$ denotes hyper-priors.
Distortion $D$ is computed as 
\begin{equation}
\begin{split}
D = \, & \textup{MSE}(\mathbf{X},\widehat{\mathbf{X}}) \\
& + \alpha \cdot \textup{MSE}\left(\sigma(\textup{B}^{(l)}(\mathbfcal{F}^{(l)})), \sigma(\textup{B}^{(l)}(\widetilde{\mathbfcal{F}}^{(l)}))\right)\\
& + \beta \cdot (1-\textup{MS-SSIM}(\mathbf{X},\widehat{\mathbf{X}})),
\end{split}
\label{eq:loss_function_1}
\end{equation}
\noindent where $\alpha$ and $\beta$ are scale factors used to achieve various trade-offs, MSE is the mean squared error, and MS-SSIM is the multi-scale structural similarity index metric~\cite{wang2003multiscale}. 

The first term in~(\ref{eq:loss_function_1}) encourages accurate reconstruction of the input image, while the third term encourages its perceptual quality. The second term is the MSE between the ground-truth feature tensor at the output of layer $l=12$ of YOLOv3 and the corresponding feature tensor derived from our base features $\mathbfcal{Y}_{\textup{base}}$. Since this term depends only on $\mathbfcal{Y}_{\textup{base}}$ and not $\mathbfcal{Y}_{\textup{enh}}$, gradients derived from it will update the model in such a way that the object detection-related information is steered towards $\mathbfcal{Y}_{\textup{base}}$. Meanwhile, both $\mathbfcal{Y}_{\textup{base}}$ and $\mathbfcal{Y}_{\textup{enh}}$ contribute to input reconstruction, so the gradients derived from the first and third term in~(\ref{eq:loss_function_1}) will distribute input reconstruction-related information across both $\mathbfcal{Y}_{\textup{base}}$ and $\mathbfcal{Y}_{\textup{enh}}$. Training of the system in Fig.~\ref{fig:overall_system} is carried out from scratch using a set of images $\mathbf{X}$ and the corresponding ground-truth feature tensors obtained at the output of layer $l=12$ of YOLOv3 for those images. 


\section{Experiments}
\label{sec:experiments}


Our model was trained on CLIC~\cite{clic_dataset} and JPEG-AI~\cite{jpeg_ai_dataset} datasets.
Images from the JPEG-AI dataset were resized to $1920\times 1080$ using the Lanczos filter. From the CLIC dataset, only images having resolutions $320\times 320$ or larger were used. Images were cropped 
using a random window with size of $256 \times 256$ during training. Ground-truth feature tensors at layer 12 of YOLOv3 were generated for all training images, to enable computing the second term in~(\ref{eq:loss_function_1}). 
During training, these tensors were cropped to $256 \times 32 \times 32$, to match the position of the random $256 \times 256$ window in the input image. 
Adam optimizer with a learning rate of $10^{-4}$ was 
used to train the network for 2M epochs on a GeForce RTX 2080 GPU with 11 GB RAM. Similarly to~\cite{minnen2018joint}, one model is trained for each $\lambda \in \{0.005, 0.01, 0.02, 0.05, 0.1, 0.2 \}$ in~(\ref{eq:RD_loss}). 


\begin{figure}[t]
    \centering
    \begin{minipage}[b]{\linewidth}
    \centering
    \includegraphics[width=\textwidth]{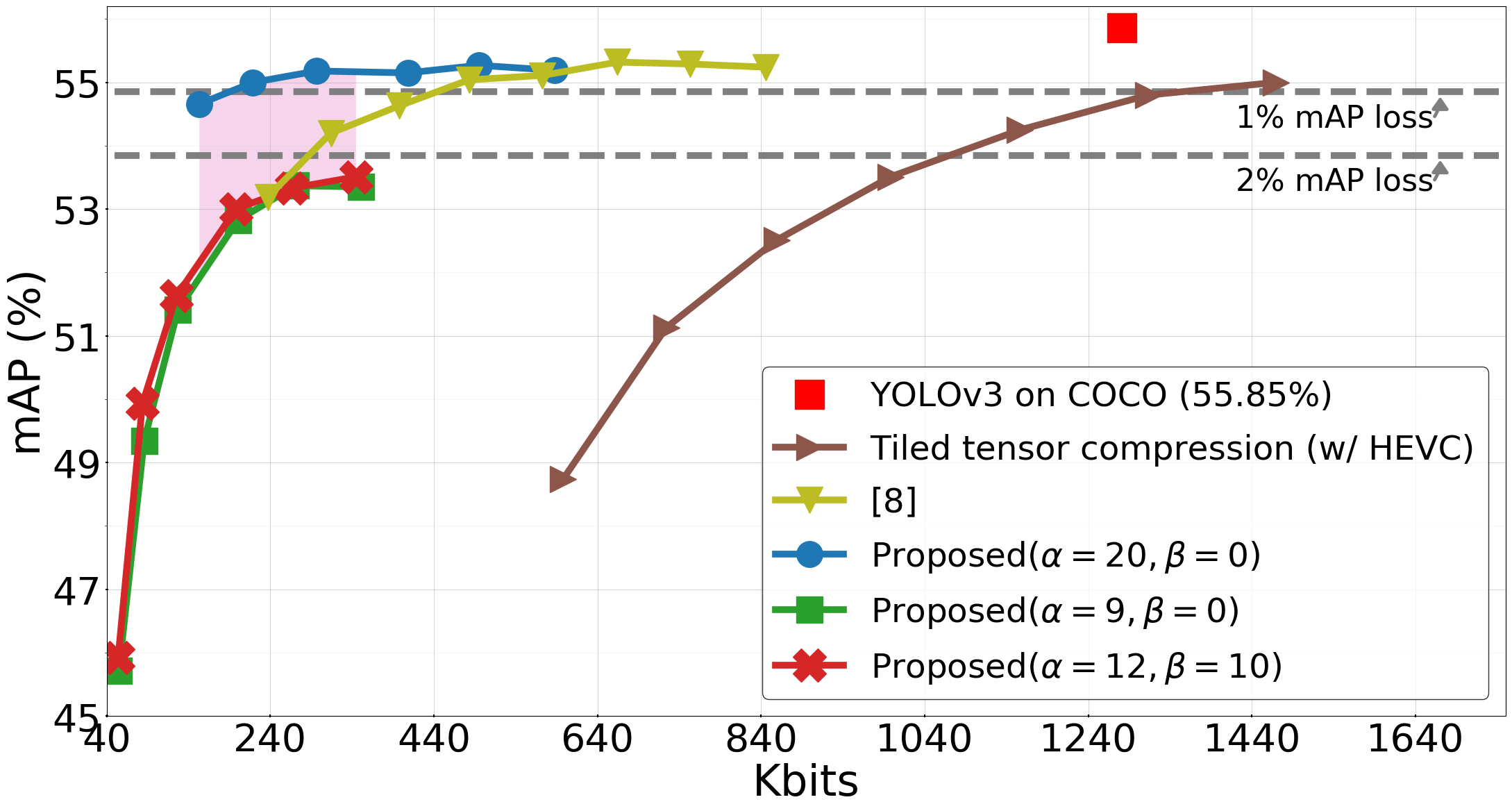}
    \end{minipage}
    \hfill
\vspace{-0.5cm}
\caption{Object detection performance of various methods.} 
\label{fig:obj_performance}
\end{figure}

Since our model supports two tasks, we compare it against relevant benchmarks for each task. For object detection, the model is evaluated on the COCO 2014 validation dataset, which includes about 5K JPEG-compressed images. The average file size of these JPEG images is around 1260 Kbits, and off-the-shelf YOLOv3 achieves the mean Average Precision (mAP) of 55.85\% on these images. This is shown as the red square in Fig.~\ref{fig:obj_performance}. When HEVC-Intra is used to encode tiled, 8-bit pre-quantized tensors from layer 12 of YOLOv3, the brown curve in Fig.~\ref{fig:obj_performance} is obtained. Our recent work~\cite{Choi_BaF_2020}, which we believe is state-of-the-art for coding YOLOv3 feature tensors, is shown as the yellow curve. To evaluate the proposed model, input images were first converted to YUV420 using \texttt{ffmpeg}, then fed to the model. Performance curves for several values of $\alpha$ and $\beta$ in~(\ref{eq:loss_function_1}) are shown as red, green, and blue in Fig.~\ref{fig:obj_performance}. Note that increasing $\alpha$ improves object detection performance, as expected from~(\ref{eq:loss_function_1}). With $\alpha=20$, outstanding performance on object detection can be achieved at very low bitrates, with less than 1\% mAP loss compared to default YOLOv3. Even with lower $\alpha$, the proposed model is competitive with~\cite{Choi_BaF_2020} at low bitrates. The shaded area in Fig.~\ref{fig:obj_performance} shows the operating range that can be achieved by varying $\alpha$.

When input reconstruction is used, our model acts as an end-to-end image codec, so we compare it against relevant benchmarks on raw YUV420 HEVC common test sequences~\cite{hevc_ctc}. 
One benchmark is HEVC (HM-16.20)~\cite{hevc_std_2019} 
with All Intra configuration~\cite{hevc_ctc}. Since the backbone of our model is based on~\cite{minnen2018joint}, we use the model from~\cite{minnen2018joint} as the second benchmark. 
To encode YUV420 input using~\cite{minnen2018joint}, chrominance channels were upsampled using nearest-neighbor interpolation and then converted to RGB using \texttt{ffmpeg}. RGB output was converted back to YUV420 using \texttt{ffmpeg}. 

Table~\ref{tbl:comp_performance} shows the average BD-Bitrate against HEVC, computed over Y-PSNR vs. bits curves. Our model with $(\alpha, \beta) = (12, 10)$ shows better performance than~\cite{minnen2018joint} on sequences in classes A, B, and C, and even better than HEVC in class A. This can be expected from~(\ref{eq:loss_function_1}), since smaller $\alpha$ together with larger $\beta$ de-emphasizes object detection performance and in turn promotes input reconstruction. Meanwhile, setting  $(\alpha, \beta)=(20,0)$ leads to a considerable loss in input reconstruction efficacy, but in turn achieves outstanding object detection performance, as seen in Fig.~\ref{fig:obj_performance}. 

Table~\ref{tbl:comp_performance_msssim} shows the average BD-Bitrate against HEVC, computed over Y-MS-SSIM vs. bits curves. It is well known that end-to-end deep model-based image codecs perform well on MS-SSIM, and this is indeed noticeable in Table~\ref{tbl:comp_performance_msssim}. Here, the model from~\cite{minnen2018joint} outperforms HEVC in all sequence classes, and our model with $(\alpha, \beta) = (12, 10)$ does so as well. Our model does better than~\cite{minnen2018joint} in class B, while in other classes,~\cite{minnen2018joint} offers better performance. It can also be noted that loss of input reconstruction efficacy of our model with $(\alpha, \beta) = (20, 0)$ is now much smaller when measured against MS-SSIM.

\begin{table}[t]
\centering
\caption{BD-Bitrate (Bits vs. Y-PSNR) vs. HM-16.20} 
\label{tbl:comp_performance}
\smallskip\noindent
\resizebox{\linewidth}{!}{%
\setlength\tabcolsep{2pt}
\renewcommand{\arraystretch}{1.2}
\begin{tabular}{@{}c|cccc@{}}
\toprule
\multirow{2}{*}{Class} & \multirow{2}{*}{\begin{tabular}[c]{@{}c@{}}Baseline\\ ~\cite{minnen2018joint}\end{tabular}} & \multicolumn{3}{c}{Proposed method} \\ \cmidrule(l){3-5} 
                       &                                                                        & ($\alpha=9, \beta=0$)         & ($\alpha=12, \beta=10$)         & ($\alpha=20, \beta=0$)         \\ \midrule \midrule
A   & 8.54\%            & -2.60\%    & \textbf{-6.93}\%    & 150.48\%   \\
B   & 43.91\%           & 9.08\%     & \textbf{3.16}\%     & 206.82\%   \\
C   & 17.98\%           & 22.17\%    & \textbf{14.57}\%    & 198.05\%   \\
D   & \textbf{17.14}\%  & 28.99\%    & 22.25\%             & 188.54\%   \\ \bottomrule
\end{tabular}}
\end{table}

\begin{table}[t]
\centering
\caption{BD-Bitrate (Bits vs. Y-MS-SSIM) vs. HM-16.20}
\label{tbl:comp_performance_msssim}
\smallskip\noindent
\resizebox{\linewidth}{!}{%
\setlength\tabcolsep{2pt}
\renewcommand{\arraystretch}{1.2}
\begin{tabular}{@{}c|cccc@{}}
\toprule
\multirow{2}{*}{Class} & \multirow{2}{*}{\begin{tabular}[c]{@{}c@{}}Baseline\\ ~\cite{minnen2018joint}\end{tabular}} & \multicolumn{3}{c}{Proposed method} \\ \cmidrule(l){3-5} 
                       &                                                                        & ($\alpha=9, \beta=0$)         & ($\alpha=12, \beta=10$)         & ($\alpha=20, \beta=0$)         \\ \midrule \midrule
A   & \textbf{-27.47}\%  & -17.11\%    & -23.72\%    & 45.64\%   \\
B   & -10.61\%  & -9.09\%     & \textbf{-16.25}\%    & 79.31\%   \\
C   & \textbf{-39.00}\%  & -3.88\%     & -10.80\%    & 80.55\%   \\
D   & \textbf{-27.84}\%  & 0.98\%      & -5.32\%     & 77.14\%   \\ \bottomrule
\end{tabular}}
\end{table}

Overall, the above results show that the proposed model can achieve comparable compression efficiency to~\cite{minnen2018joint} when used as and end-to-end image codec, and can be better than HEVC when the reconstruction quality is measured using MS-SSIM. On top of that, our model offers scalability to perform object detection from a subset of the latent space, which neither~\cite{minnen2018joint} nor HEVC (nor any other codec, to our knowledge) is currently able to offer. We therefore believe it will be a useful contribution to future research and standardization activities in this area.

\section{Conclusions}
\label{sec:conclusion}
We introduced latent-space scalability for multi-task collaborative intelligence, and tested it on a system that supports object detection and input reconstruction. A part of the latent space is dedicated to object detection, while the whole latent space is used for input reconstruction. Appropriately chosen loss terms allow for steering relevant information to different portions of the latent space as the model is trained. 
By varying the scaling factors of various loss terms, different trade-offs between the two tasks were demonstrated and compared with 
the relevant benchmarks.


\bibliographystyle{IEEEbib-abbrev} 
\small
\bibliography{ref}

\end{document}